\documentclass[10pt,a4paper]{article}
\usepackage{amsmath,amssymb,amsfonts}
\usepackage{graphicx}
\usepackage{hyperref}
\usepackage[margin=0.75in]{geometry}
\usepackage{float}
\usepackage{bm}
\usepackage{booktabs}
\usepackage{tabularx}
\usepackage{subcaption}

\usepackage{setspace}
\usepackage{titlesec}
\titlespacing*{\section}{0pt}{8pt}{4pt}
\titlespacing*{\subsection}{0pt}{6pt}{3pt}

\AtBeginDocument{
  \setlength{\abovedisplayskip}{4pt}
  \setlength{\belowdisplayskip}{4pt}
  \setlength{\abovedisplayshortskip}{2pt}
  \setlength{\belowdisplayshortskip}{2pt}
}

\let\oldbibliography\thebibliography
\renewcommand{\thebibliography}[1]{%
  \oldbibliography{#1}%
  \setlength{\itemsep}{0pt}%
  \setlength{\parskip}{0pt}%
}

\title{Measuring Primitive Accumulation: An Information-Theoretic Approach to Capitalist Enclosure in PIK2, Indonesia}

\author{
Sandy Hardian Susanto Herho$^{1,2,*}$,
Alfita Puspa Handayani$^{1,3}$,
Karina Aprilia Sujatmiko$^{4}$,\\
Faruq Khadami$^{4}$, Iwan Pramesti Anwar$^{4}$, Rusmawan Suwarman$^{5}$,\\
Dasapta Erwin Irawan$^{2}$, and Deny Juanda Puradimaja$^{2}$
}

\date{}

\begin{document}
\maketitle

\begin{center}
\small
$^{1}$Center for Agrarian Studies, Bandung Institute of Technology, Bandung, West Java 40132, Indonesia\\
$^{2}$Applied Geology Research Group, Bandung Institute of Technology, Bandung, West Java 40132, Indonesia\\
$^{3}$Spatial Systems and Cadaster Research Group, Bandung Institute of Technology, Bandung, West Java 40132, Indonesia\\
$^{4}$Applied and Environmental Oceanography Research Group, Bandung Institute of Technology, Bandung, West Java 40132, Indonesia\\
$^{5}$Atmospheric Science Research Group, Bandung Institute of Technology, Bandung, West Java 40132, Indonesia\\
$^{*}$Correspondence: sandyherho@itb.ac.id
\end{center}

\begin{abstract}
\noindent
Large-scale land enclosure for speculative mega-development constitutes a non-equilibrium spatial process whose velocity, topology, and irreversibility remain poorly quantified. We study the Pantai Indah Kapuk 2 (PIK2) coastal mega-development north of Jakarta, Indonesia, using eight years (2017--2024) of Sentinel-2 land-use/land-cover (LULC) data at 10-meter resolution. The landscape is projected onto a Marxian probability simplex partitioning terrestrial pixels into Commons, Agrarian, and Capital fractions. Fisher-Rao (FR) geodesic distances on this simplex identify a transformation pulse of $0.405$~rad/yr during 2019--2020, coinciding with major construction activity. Absorbing Markov chain analysis yields expected absorption times into the built environment of $46.0$~years for cropland and $38.1$~years for tree cover, with a pooled built-area self-retention rate of $96.4\%$. Percolation analysis reveals that a giant connected component containing $89$--$95\%$ of all built pixels persists at occupation probabilities $p \in [0.096, 0.162]$, far below the random percolation threshold $p_c \approx 0.593$, indicating planned rather than stochastic spatial growth. The box-counting fractal dimension of the urban boundary increases from $d_f = 1.316$ to $1.397$, consistent with increasingly irregular frontier expansion. These results suggest that information-geometric and statistical-mechanical tools can characterize the kinematic and topological signatures of capitalist spatial accumulation with quantitative precision.
\end{abstract}


\section{Introduction}
\label{sec:intro}

The spatial expansion of cities and mega-developments can be understood as a non-equilibrium dynamical process on a discrete lattice, subject to external driving forces such as capital investment, state regulation, and infrastructure planning that push the system far from any stationary configuration \cite{Batty2007, Makse1995}. In the language of statistical mechanics, such processes involve the irreversible conversion of lattice sites from one state (agricultural, natural) to another (built, commodified), with transition rates that fluctuate in response to political and economic shocks. While urban growth has been studied through percolation models \cite{Makse1995, Rozenfeld2008} and fractal geometry \cite{Batty2007, Mandelbrot1982}, these approaches are seldom integrated with the structural categories of political economy that explain \emph{why} particular land conversions occur.

Within critical geography, the concept of \emph{primitive accumulation} (PA) describes the ongoing process by which non-capitalist spatial frontiers, including commons and peasant agriculture, are enclosed and converted into commodified real estate \cite{Marx1976, Harvey2003, DeAngelis2001}. PA is not a historical relic confined to early industrialization; rather, it operates as a recurring structural mechanism through which capital resolves crises of overaccumulation via what Harvey \cite{Harvey2003} terms the "spatial fix" and what Smith \cite{Smith2008} analyzes as the production of uneven development. In the contemporary era of planetary urbanization \cite{Brenner2014, Luxemburg1913}, this enclosure is concentrated in the Global South, where state-backed mega-developments rapidly subsume agrarian landscapes into speculative real estate \cite{Glassman2006}.

This paper develops a quantitative framework, rooted in information geometry (IG) \cite{Amari2016}, Markov chain theory \cite{Kemeny1976}, and percolation theory \cite{Stauffer1992}, to measure the rate, directionality, and spatial topology of such enclosure. We apply this framework to the Pantai Indah Kapuk 2 (PIK2) mega-development on the northern coast of Jakarta, Indonesia. Jakarta, the capital of Indonesia and a metropolitan area of over 30~million inhabitants, sits on a low-lying alluvial plain along the Java Sea. The city's northern coast has experienced severe land subsidence of up to 25~cm/yr in some areas and recurrent tidal flooding \cite{Abidin2011, Erkens2015, Garschagen2018}. Despite these hazards, the coastal zone has become a site of intensive speculative development.

PIK2, developed by private consortia on reclaimed and expropriated coastal land approximately 20~km northwest of central Jakarta, represents one of the largest and most rapid spatial transformations in Southeast Asia \cite{Colven2017}. The project encompasses residential townships, commercial districts, and extensive road infrastructure built over what was, until recently, a mosaic of fish ponds, mangrove remnants, and smallholder agriculture.

The political economy of PIK2 is characterized by intense regulatory volatility. In early 2024, the administration of President Joko Widodo designated PIK2 as a National Strategic Project (\emph{Proyek Strategis Nasional}, PSN), granting developers accelerated land acquisition powers \cite{Kemenko2024PSN}. This designation followed, rather than preceded, a period of aggressive land clearance that is visible in the satellite record. The resulting agrarian conflicts and public scrutiny prompted the subsequent administration of President Prabowo Subianto to initiate a formal review and partial suspension of PSN land acquisition privileges for PIK2 \cite{Alfiana2026}. This sequence underscores that capitalist enclosure is not a smooth, deterministic process but a contested, non-stationary one shaped by class conflict, legislative shocks, and political friction.

Our contributions are as follows. First, we project high-resolution satellite land-use data onto a probability simplex and use the Fisher-Rao (FR) geodesic distance, the unique Riemannian metric invariant under sufficient statistics, to measure the "velocity" of landscape transformation. Second, we formulate the landscape dynamics as an absorbing Markov chain with the built environment as the absorbing state, yielding exact expected absorption times for each land-use class. Third, we apply site percolation analysis to characterize the spatial connectivity and fractal morphology of the expanding built environment. Together, these tools provide a statistical-mechanical diagnostic for the kinematic, stochastic, and topological dimensions of spatial enclosure.

\section{Study area and data}
\label{sec:data}

\begin{figure}[H]
    \centering
    \includegraphics[width=\linewidth]{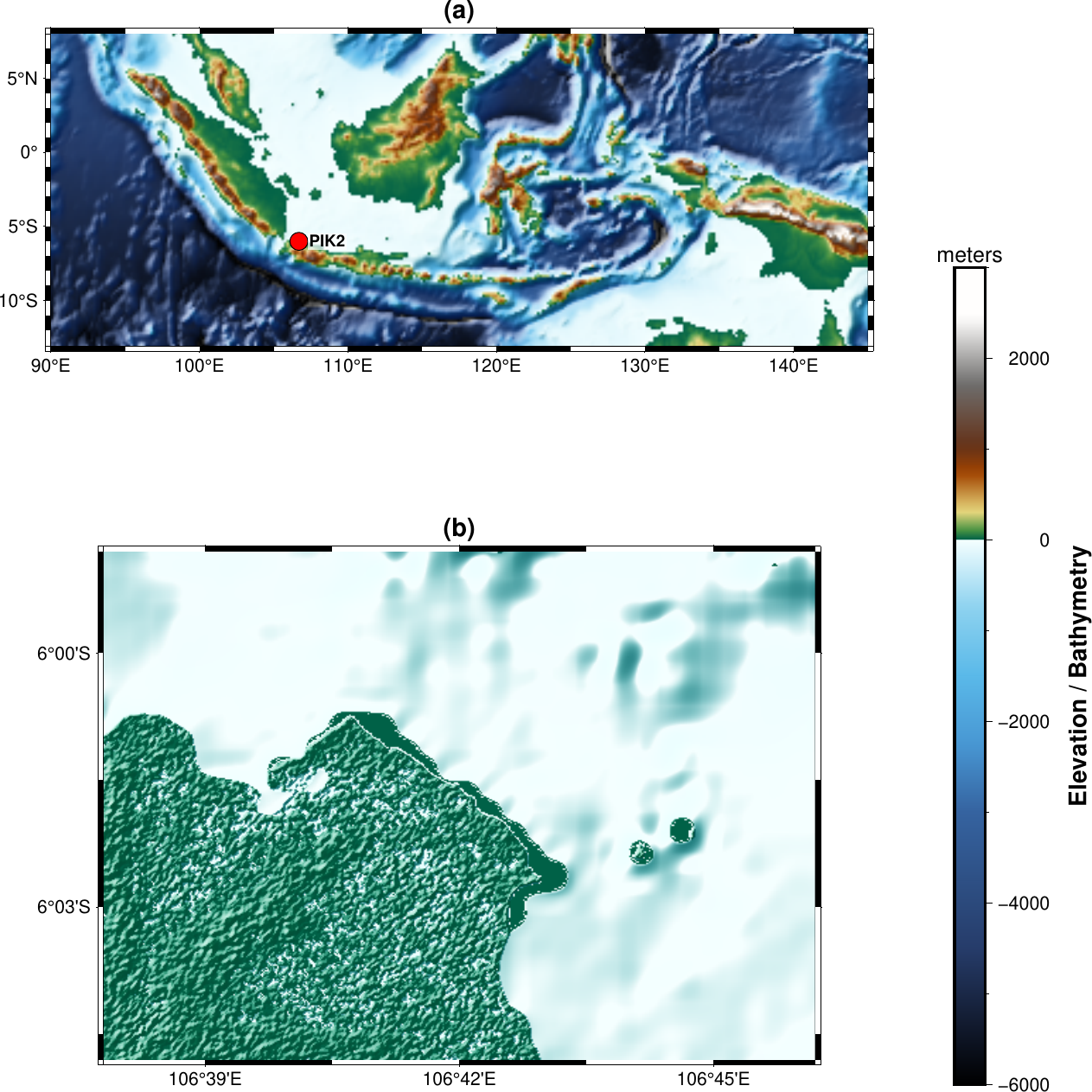}
    \caption{Geographic context of the study area. (a) Location of PIK2 within the Indonesian archipelago. (b) High-resolution (1 arc-second, ${\sim}30$~m) bathymetric and topographic relief of the PIK2 domain, showing the low-lying coastal substrate. Data derived from the Shuttle Radar Topography Mission (SRTM) \cite{Farr2007SRTM}, rendered via PyGMT \cite{Wessel2019GMT}.}
    \label{fig:study_area}
\end{figure}

The study domain is bounded by longitude $106.63^{\circ}\mathrm{E}$ to $106.77^{\circ}\mathrm{E}$ and latitude $6.08^{\circ}\mathrm{S}$ to $5.98^{\circ}\mathrm{S}$, encompassing the PIK2 mega-development and its surrounding coastal and agrarian hinterland on the northern shore of Jakarta, Java, Indonesia (Figure~\ref{fig:study_area}). The domain covers $173.4$~km$^2$, of which approximately $70\%$ is shallow ocean. Analysis of 1-arcsecond bathymetric and topographic data reveals that the marine portion (126{,}223 pixels) has a mean depth of $-6.68$~m and a maximum depth of $-22.32$~m, while the terrestrial portion (65{,}913 pixels) has a mean elevation of only $2.60$~m and a maximum of $14.00$~m. This extremely low-lying topography makes the area both attractive for reclamation-based development and ecologically vulnerable to subsidence and flooding.

We use the Esri Sentinel-2 10-Meter Annual Land Use and Land Cover (LULC) time series \cite{Karra2021Esri}, which provides a globally consistent, deep-learning-derived classification at $10$~m spatial resolution. The temporal domain spans eight annual composites from 2017 to 2024, denoted $\mathcal{T} = \{2017, 2018, \ldots, 2024\}$ with $|\mathcal{T}| = 8$. The raw GeoTIFF rasters are reprojected onto a uniform World Geodetic System 1984 (WGS-84) grid using the Rasterio and PyProj libraries \cite{Gillies2013Rasterio, PROJContributors2021} and compiled into a compressed Network Common Data Form version~4 (NetCDF4) tensor \cite{Rew1990NetCDF}.

The resulting spatial lattice $\Lambda$ consists of $1{,}113$ rows $\times$ $1{,}558$ columns $= 1{,}734{,}054$ pixels, with zero no-data and zero cloud-contaminated pixels across all years. The classification assigns each pixel to one of $K = 7$ substantive classes after filtering: Water ($k{=}1$), Trees ($k{=}2$), Flooded Vegetation ($k{=}4$), Crops ($k{=}5$), Built Area ($k{=}7$), Bare Ground ($k{=}8$), and Rangeland ($k{=}11$). Snow/Ice and Clouds are absent in this tropical, low-elevation domain.

For the macrostructural analysis, we exclude the marine boundary (Water, which constitutes ${\sim}70\%$ of the domain and is not subject to land enclosure) and aggregate the remaining terrestrial classes into three categories motivated by the structural categories of political economy \cite{Harvey2003, Marx1976}. "Commons" groups Trees, Flooded Vegetation, and Rangeland, representing areas outside the direct circuit of commodification and historically utilized through customary rights. "Agrarian" corresponds strictly to Crops, representing spaces of petty commodity production and smallholder agriculture. "Capital" groups Built Area and Bare Ground, representing the commodified built environment together with its construction-phase precursor. Bare Ground is included within Capital rather than treated as a natural state because, in coastal reclamation and mega-development contexts, exposed bare soil typically represents recently cleared or graded land awaiting construction \cite{Lambin2001}. The empirical validity of this assignment is examined through the Markov transition analysis.

\section{Methods}
\label{sec:methods}

\subsection{Empirical probability distributions on the lattice}
\label{subsec:pmf}

Let $\Lambda = \{\mathbf{x}_i\}_{i=1}^{|\Lambda|}$ denote the set of spatial sites (pixel centroids) on the regular two-dimensional lattice, where each site $\mathbf{x} \in \mathbb{R}^2$ corresponds to a $10 \times 10$~m$^2$ area \cite{Cressie1993, Lambin2001}. Let $\mathcal{K} = \{1, 2, \ldots, K\}$ with $K = 7$ be the set of LULC classes. The landscape state at time $t \in \mathcal{T}$ is described by a discrete random field $C \colon \Lambda \times \mathcal{T} \to \mathcal{K}$, mapping each site at each time to a class label.

The empirical probability mass function (PMF) of class $k$ at time $t$ is
\begin{equation}
    p_k(t) = \frac{1}{N(t)} \sum_{\mathbf{x} \in \Lambda} \mathbb{I}_{\{C(\mathbf{x},t) = k\}},
    \label{eq:pmf}
\end{equation}
where $\mathbb{I}_{\{\cdot\}}$ is the indicator function (equal to 1 when the condition holds and 0 otherwise) and $N(t) = |\Lambda| = 1{,}734{,}054$ is the number of valid pixels, constant across years in this dataset.

The resulting state vector $\mathbf{p}(t) = [p_1(t),\, p_2(t),\, \ldots,\, p_K(t)]^\top$ resides on the standard $(K{-}1)$-dimensional probability simplex,
\begin{equation}
    \Delta^{K-1} = \left\{ \mathbf{p} \in \mathbb{R}^K \;\middle|\; p_k \geq 0\ \forall k \in \mathcal{K},\; \sum_{k=1}^K p_k = 1 \right\}.
\label{eq:simplex}
\end{equation}
For the ternary Marxian aggregation, we define a reduced state vector $\mathbf{q}(t) = [q_{\mathrm{com}}(t),\, q_{\mathrm{agr}}(t),\, q_{\mathrm{cap}}(t)]^\top$ on the 2-simplex $\Delta^2$, where each component is computed by summing the relevant fine-grained class probabilities over land pixels only (excluding Water). All computations use NumPy \cite{Harris2020NumPy}, SciPy \cite{Virtanen2020SciPy}, and Matplotlib \cite{Hunter2007Matplotlib}.

\subsection{Information-theoretic measures}
\label{subsec:info_theory}

We quantify landscape heterogeneity and transformation velocity using four information-theoretic quantities, all computed on the full $K{=}7$ class distribution $\mathbf{p}(t)$. The Shannon entropy \cite{Shannon1948, Cover2006} of the landscape at time $t$ is
\begin{equation}
    H(\mathbf{p}(t)) = -\sum_{k=1}^{K} p_k(t) \ln p_k(t),
    \label{eq:shannon}
\end{equation}
measured in nats, where the convention $0 \ln 0 = 0$ applies. $H$ attains its maximum value of $\ln K = \ln 7 \approx 1.946$~nats when all classes are equiprobable, and its minimum of zero when a single class dominates completely.

To probe the multi-scale structure of the class distribution, we also compute the R\'{e}nyi entropy of order $\alpha$ \cite{Renyi1961},
\begin{equation}
    H_\alpha(\mathbf{p}(t)) = \frac{1}{1-\alpha} \ln\left(\sum_{k=1}^{K} p_k(t)^\alpha\right),
    \label{eq:renyi}
\end{equation}
where $\alpha \geq 0$ is a real-valued parameter controlling sensitivity to rare versus dominant classes. In the limits, $H_0 = \ln|\mathrm{supp}(\mathbf{p})|$ counts the support size, $H_1 \equiv \lim_{\alpha \to 1} H_\alpha = H$ recovers the Shannon entropy, and $H_\infty = -\ln(\max_k p_k)$ is sensitive only to the most probable class. The monotonic tendency of the entropy time series $\{H(t)\}_{t \in \mathcal{T}}$ is assessed using the non-parametric Mann-Kendall (MK) test, computing the Kendall rank correlation $\tau$ and its associated $p$-value under the null hypothesis of no trend.

The directional information gain between consecutive years is measured by the Kullback-Leibler divergence (KLD) \cite{Kullback1951, Cover2006},
\begin{equation}
    D_{\mathrm{KL}}\bigl(\mathbf{p}(t{+}1) \,\|\, \mathbf{p}(t)\bigr) = \sum_{k=1}^{K} p_k(t{+}1) \ln\frac{p_k(t{+}1)}{p_k(t)},
    \label{eq:kl}
\end{equation}
defined for all $k$ where $p_k(t) > 0$. KLD is non-negative, equals zero if and only if $\mathbf{p}(t{+}1) = \mathbf{p}(t)$, and is asymmetric. Because it is neither symmetric nor satisfies the triangle inequality, it cannot serve as a proper metric on $\Delta^{K-1}$.

The FR distance \cite{Amari2016} fills this role, providing the unique Riemannian metric on the probability simplex that is invariant under sufficient statistics. For two distributions $\mathbf{p}$ and $\mathbf{q}$ on $\Delta^{K-1}$, the FR distance is computed via the Bhattacharyya coefficient (BC),
\begin{equation}
    \mathrm{BC}(\mathbf{p}, \mathbf{q}) = \sum_{k=1}^{K} \sqrt{p_k \, q_k} \;\in\; [0, 1],
    \label{eq:bc}
\end{equation}
which equals 1 when $\mathbf{p} = \mathbf{q}$ and 0 when the supports are disjoint. The FR geodesic distance is then
\begin{equation}
    d_{\mathrm{FR}}(\mathbf{p}, \mathbf{q}) = 2\arccos\bigl(\mathrm{BC}(\mathbf{p}, \mathbf{q})\bigr),
    \label{eq:fisher_rao}
\end{equation}
yielding a value in $[0, \pi]$~rad that is symmetric and satisfies the triangle inequality.

To identify periods of anomalously rapid transformation, we define a \emph{transformation pulse} at time $t$ as any transition satisfying
\begin{equation}
    d_{\mathrm{FR}}\bigl(\mathbf{p}(t), \mathbf{p}(t{+}1)\bigr) > \bar{d}_{\mathrm{FR}} + s_{\mathrm{FR}},
    \label{eq:pulse}
\end{equation}
where $\bar{d}_{\mathrm{FR}}$ and $s_{\mathrm{FR}}$ denote the sample mean and standard deviation (with Bessel's correction) of the $|\mathcal{T}|{-}1 = 7$ consecutive FR distances. We additionally compute the FR distance on the reduced 3-simplex $\Delta^2$ using $\mathbf{q}(t)$ in place of $\mathbf{p}(t)$. The cumulative arc length $L(t) = \sum_{\tau < t} d_{\mathrm{FR}}(\mathbf{q}(\tau), \mathbf{q}(\tau{+}1))$, the direct geodesic displacement $D = d_{\mathrm{FR}}(\mathbf{q}(2017), \mathbf{q}(2024))$, and the sinuosity $\sigma = L/D$ characterize the non-linearity of the simplex trajectory.

\subsection{Markov chain formulation}
\label{subsec:markov}

The pixel-wise temporal dynamics are modeled as a discrete-time Markov chain on the state space $\mathcal{K}$ \cite{Kemeny1976, Norris1997}. For each consecutive pair of years $(t, t{+}1)$, we construct a $K \times K$ count matrix $\mathbf{C}(t)$, where element $C_{ij}(t)$ records the number of pixels that transition from class $i$ at time $t$ to class $j$ at time $t{+}1$. The empirical transition probability is obtained by row normalization,
\begin{equation}
    p_{ij}(t) = \frac{C_{ij}(t)}{n_i(t)}, \quad \text{where} \quad n_i(t) = \sum_{j=1}^{K} C_{ij}(t)
    \label{eq:trans_prob}
\end{equation}
is the total number of pixels originating in class $i$. The associated asymptotic standard error (SE) is $\mathrm{SE}(p_{ij}) = \sqrt{p_{ij}(1 - p_{ij})/n_i}$. A pooled transition matrix $\mathbf{P}^{(\mathrm{pool})}$ is obtained by summing all seven count matrices element-wise, $\mathbf{C}^{(\mathrm{pool})} = \sum_{t} \mathbf{C}(t)$, and then row-normalizing.

To quantify the temporal horizon of land conversion, we reformulate the system as an absorbing Markov chain \cite{Kemeny1976} by designating Built Area (class $k{=}7$, hereafter $B$) as the sole absorbing state, satisfying $p_{BB} = 1$ and $p_{Bj} = 0$ for all $j \neq B$. Let $\mathcal{K}_T = \mathcal{K} \setminus \{B\}$ denote the set of $K_T = K - 1 = 6$ transient states, and let $\mathbf{Q}$ be the $K_T \times K_T$ sub-matrix of transition probabilities among them. The fundamental matrix is
\begin{equation}
    \mathbf{N} = (\mathbf{I}_{K_T} - \mathbf{Q})^{-1},
    \label{eq:fundamental}
\end{equation}
where $\mathbf{I}_{K_T}$ is the $K_T \times K_T$ identity matrix. Element $N_{ij}$ gives the expected number of time steps the chain spends in transient state $j$ given that it starts in state $i$. The expected absorption time from each transient state is the row sum
\begin{equation}
    \mathbb{E}[T_i] = \sum_{j \in \mathcal{K}_T} N_{ij} = (\mathbf{N}\mathbf{1})_i,
    \label{eq:absorption_time}
\end{equation}
where $\mathbf{1}$ is a column vector of ones of length $K_T$. The variance of the absorption time is
\begin{equation}
    \mathrm{Var}[T_i] = \bigl[(2\mathbf{N} - \mathbf{I}_{K_T})\,\mathbb{E}[\mathbf{T}]\bigr]_i - \mathbb{E}[T_i]^2,
    \label{eq:var_absorption}
\end{equation}
and $\mathrm{SD}[T_i] = \sqrt{\mathrm{Var}[T_i]}$ characterizes the intrinsic spread of the first-passage-time distribution for individual pixel trajectories.

The stationary distribution $\boldsymbol{\pi}$ of the empirical (non-forced) pooled matrix satisfies $\boldsymbol{\pi}^\top \mathbf{P}^{(\mathrm{pool})} = \boldsymbol{\pi}^\top$ with $\sum_k \pi_k = 1$, and is obtained as the left eigenvector corresponding to eigenvalue $\lambda_1 = 1$. We test the null hypothesis that the transition matrix is constant over time using the log-likelihood ratio $G$-test \cite{Sokal2012},
\begin{equation}
    G = 2 \sum_{t} \sum_{i=1}^{K} \sum_{j=1}^{K} C_{ij}^{(t)} \ln\frac{p_{ij}^{(t)}}{p_{ij}^{(\mathrm{pool})}},
    \label{eq:gtest}
\end{equation}
where the sum over $t$ runs over all seven transition periods. Under the null, $G$ is asymptotically $\chi^2$-distributed. The period-specific deviation from the pooled matrix is further quantified by the Frobenius norm $\|\mathbf{P}(t) - \mathbf{P}^{(\mathrm{pool})}\|_F$.

\subsection{Percolation analysis}
\label{subsec:percolation}

We apply site percolation theory \cite{Stauffer1992, Grimmett1999} to the built environment. Each pixel classified as Built Area at time $t$ is treated as an occupied site on the two-dimensional lattice $\Lambda \subset \mathbb{Z}^2$. Two occupied sites are assigned to the same connected cluster $\mathcal{C}$ if they share an orthogonal edge, following the standard von Neumann 4-connectivity rule. Connected component labeling is performed using the \texttt{scipy.ndimage.label} algorithm \cite{Virtanen2020SciPy}.

Let $S_{\mathrm{built}}(t)$ denote the total number of occupied (built) pixels at time $t$, and let $|\Lambda|$ denote the total number of lattice sites. The global occupation probability is defined as
\begin{equation}
    p(t) = \frac{S_{\mathrm{built}}(t)}{|\Lambda|}.
\label{eq:occ}
\end{equation}
Let $N_{\mathrm{cl}}(t)$ denote the number of distinct connected clusters and let $S_{\max}(t)$ denote the size (in pixels) of the largest among them. The macroscopic spatial connectivity of the system is evaluated via the order parameter
\begin{equation}
    \Omega(t) = \frac{S_{\max}(t)}{S_{\mathrm{built}}(t)},
    \label{eq:order_param}
\end{equation}
which measures the fraction of all built pixels contained within the single largest cluster. For uncorrelated random site percolation on the infinite square lattice, a spanning (giant) component emerges only above the critical occupation threshold $p_c \approx 0.5927$ \cite{Stauffer1992}. Values of $\Omega$ close to unity at occupation probabilities $p \ll p_c$ would therefore indicate the presence of strong spatial correlations among occupied sites.

The morphological complexity of the largest cluster's boundary is quantified by the box-counting fractal dimension $d_f$ \cite{Mandelbrot1982}. Let $\partial \mathcal{C}$ denote the topological boundary of the largest connected component, obtained by subtracting the morphologically eroded interior (using a $3 \times 3$ structuring element) from the boolean cluster mask. The boundary $\partial \mathcal{C}$ is then covered with a sequence of square boxes of geometrically decreasing side length $\epsilon$, and the number of non-empty boxes $\mathcal{N}(\epsilon)$ is recorded at each scale. The fractal dimension is extracted as the slope of the power-law scaling relation
\begin{equation}
    \ln \mathcal{N}(\epsilon) = -d_f \ln \epsilon + c,
    \label{eq:fractal}
\end{equation}
where $c$ is an intercept constant. The slope $d_f$ and its coefficient of determination $R^2$ are estimated via ordinary least-squares (OLS) regression \cite{Virtanen2020SciPy}. Reference universality classes for two-dimensional boundary morphology include compact growth ($d_f \to 1.0$), Eden-model growth ($d_f \approx 1.50$), and Diffusion-Limited Aggregation (DLA) boundaries ($d_f \approx 1.71$--$1.75$) \cite{Witten1981}. The complementary cumulative distribution function (CCDF) of cluster sizes, $P(S \geq s)$, is also computed for each year. At the critical point of random percolation, the cluster-size distribution follows a power law with Fisher exponent $\tau \approx 2.05$ \cite{Stauffer1992}. Deviations from this scaling in the empirical data provide additional information about the degree and nature of spatial correlations in the system.

\section{Results}
\label{sec:results}

\begin{figure}[H]
    \centering
    \includegraphics[width=\linewidth]{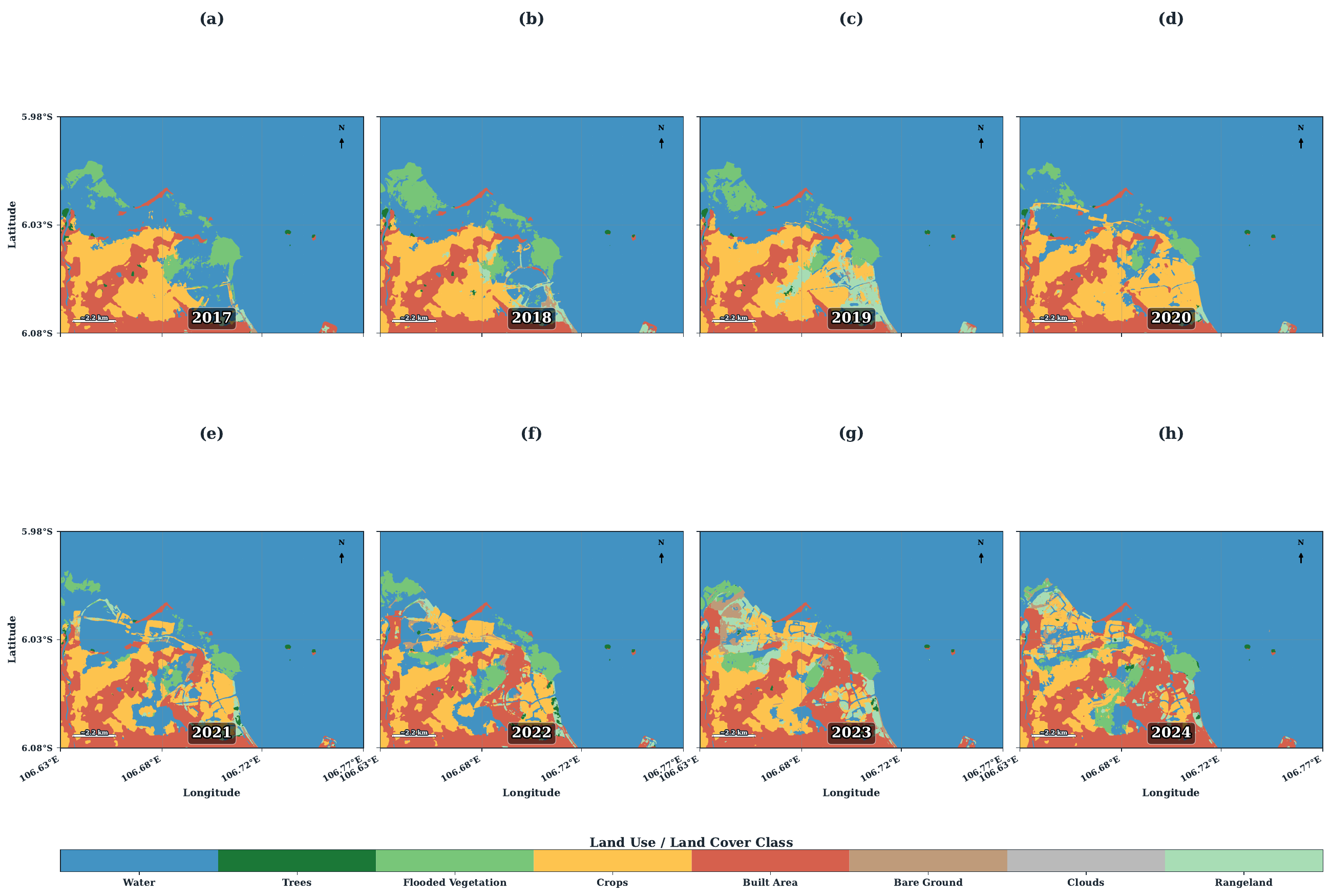}
    \caption{Annual LULC classification of the PIK2 domain (2017--2024), showing the progressive expansion of Built Area (red) at the expense of Crops (yellow) and natural vegetation (green).}
    \label{fig:lulc_map}
\end{figure}

Figure~\ref{fig:lulc_map} displays the spatial configuration of the PIK2 domain across all eight years. The lattice remains invariant at $1{,}734{,}054$ pixels ($173.4$~km$^2$), with no cloud contamination or missing data in any year. Water is the dominant class throughout, declining from $75.4\%$ (2017) to $68.8\%$ (2024), consistent with ongoing coastal reclamation. The terrestrial landscape undergoes a rapid structural shift. Built Area expands nearly monotonically from $166{,}405$ pixels ($9.60\%$, $16.6$~km$^2$) in 2017 to $280{,}326$ pixels ($16.17\%$, $28.0$~km$^2$) in 2024, a net gain of $+6.57$ percentage points (pp). Crops peak at $13.90\%$ in 2020 before declining to $9.15\%$ by 2024. Tree cover remains below $0.34\%$ throughout the study period, indicating that large-scale deforestation in this region had already been completed before 2017.

\begin{figure}[H]
    \centering
    \includegraphics[width=\linewidth]{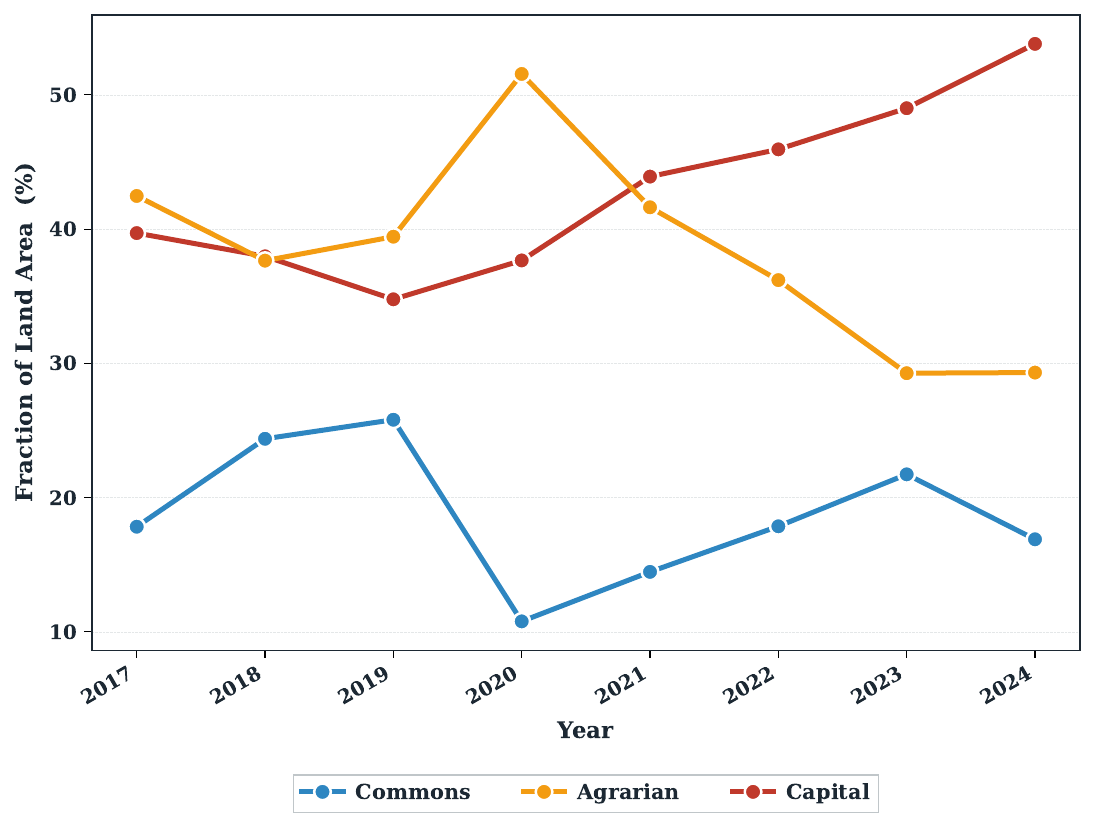}
    \caption{Trajectory of the terrestrial landscape on the Marxian ternary simplex (land pixels only, ocean excluded). Capital surpasses Agrarian between 2020 and 2021.}
    \label{fig:simplex}
\end{figure}

Projecting the land-only pixels onto the Marxian ternary simplex (Figure~\ref{fig:simplex}) reveals the macrostructural trajectory of enclosure. In 2017, Agrarian dominated terrestrial land at $42.47\%$, followed by Capital at $39.70\%$ and Commons at $17.83\%$. The trajectory is non-monotonic: Agrarian spikes to $51.56\%$ in 2020, possibly reflecting temporary reclassification during pandemic-era construction slowdowns, before declining to $29.31\%$ by 2024. Capital increases from $39.70\%$ to $53.80\%$ over the same period ($\Delta = +14.1$~pp, $+36\%$ relative), while Commons declines modestly from $17.83\%$ to $16.89\%$ ($\Delta = -0.9$~pp). The dominant secular transfer is from Agrarian to Capital: cropland loses $13.2$~pp while Capital gains $14.1$~pp, a nearly one-to-one replacement. This pattern is consistent with the Marxian account of PA, where productive agricultural land is directly converted into commodified real estate \cite{Harvey2003, DeAngelis2001}. The Capital-Agrarian crossover occurs around 2020--2021, when commodified land area first exceeds that of smallholder agriculture.

An internal decomposition of Capital is informative. The ratio of completed Built Area to total Capital (Built $+$ Bare Ground) remains above $0.95$ in most years, indicating that the vast majority of capitalized land is fully developed. A notable exception occurs in 2023, when this ratio drops to $0.901$: nearly $10\%$ of all Capital is in an active state of land clearance and bare-earth preparation. This spike in the Bare Ground fraction (from $2.40\%$ to $4.87\%$ of land) directly precedes the March 2024 PSN designation \cite{Kemenko2024PSN}, suggesting that speculative land clearance was already underway before the formal regulatory authorization was issued.

The FR distances on the 3-simplex yield a cumulative arc length of $L = 1.251$~rad over the seven transitions, compared to a direct displacement of $D = 0.303$~rad. The resulting sinuosity $\sigma = L/D = 4.12$ indicates a highly non-linear path: the landscape zigzags through seasonal and construction-related fluctuations while drifting secularly toward the Capital vertex. Peak velocity occurs during 2019--2020 ($d_{\mathrm{FR}} = 0.405$~rad/yr), consistent with major PIK2 construction activity.

\begin{figure}[H]
    \centering
    \includegraphics[width=\linewidth]{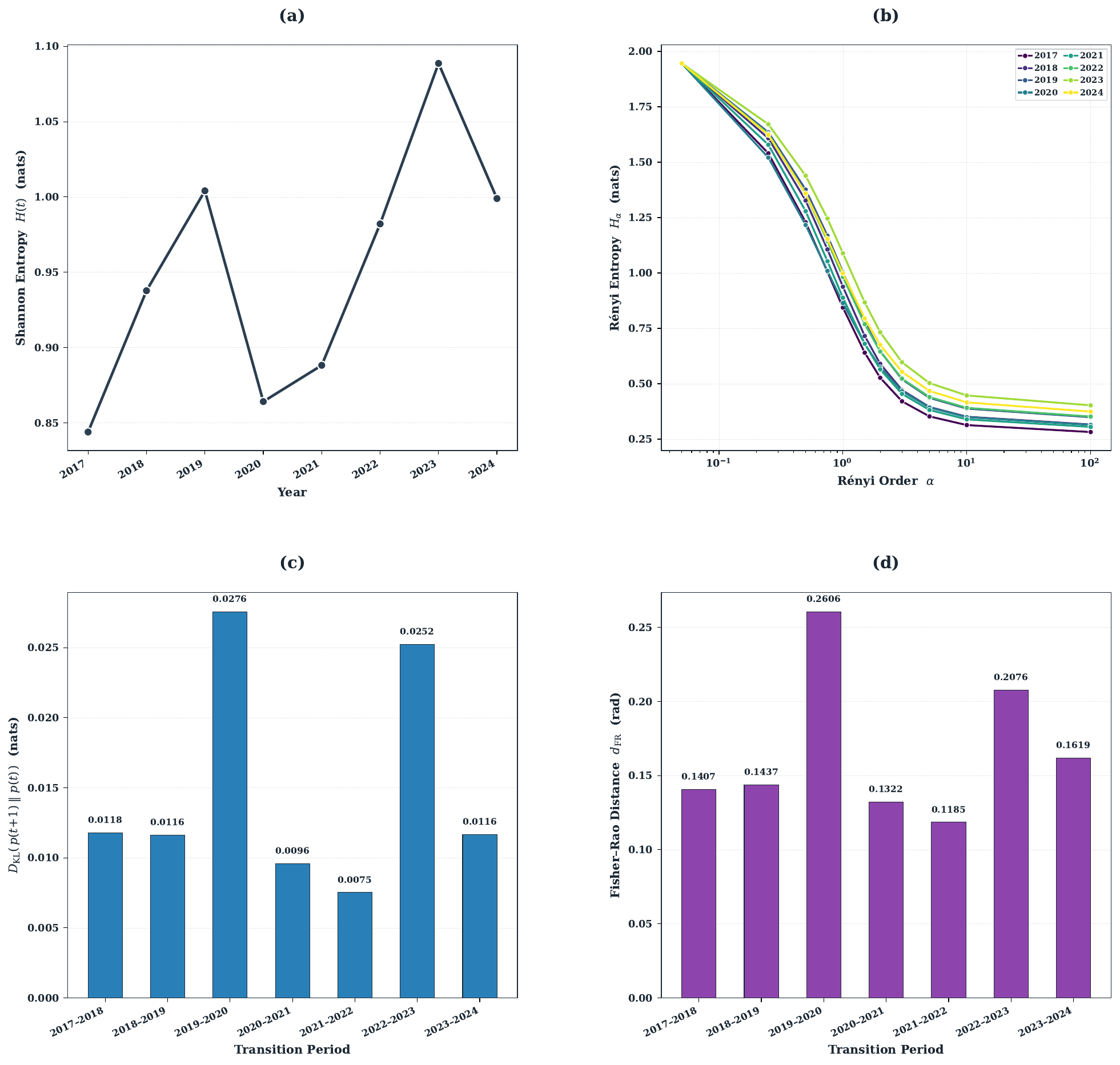}
    \caption{Information-theoretic analysis of the 7-class LULC distribution. (a) Shannon entropy $H(t)$. (b) R\'{e}nyi entropy spectrum $H_\alpha(t)$. (c) KLD $D_{\mathrm{KL}}(\mathbf{p}(t{+}1)\|\mathbf{p}(t))$. (d) FR geodesic distance $d_{\mathrm{FR}}(\mathbf{p}(t), \mathbf{p}(t{+}1))$.}
    \label{fig:entropy}
\end{figure}

Figure~\ref{fig:entropy} summarizes the information-theoretic analysis of the full 7-class distribution $\mathbf{p}(t)$. Shannon entropy $H(t)$ ranges from $0.844$~nats (2017) to $1.089$~nats (2023), with a net increase of $+0.155$~nats over the study period. The MK test yields $\tau = 0.50$ with $p = 0.108$, indicating a positive trend that does not reach statistical significance at the $\alpha = 0.05$ level. The entropy increase may appear counterintuitive if one expects enclosure to reduce landscape diversity. However, because the domain is ${\sim}70\%$ ocean, the growth of Built Area from $9.6\%$ to $16.2\%$ diversifies the distribution away from the Water-dominated baseline, temporarily raising $H(t)$. A reversal (entropy decline) would be expected only once Built Area surpasses the dominant class. This pattern is consistent with a pre-enclosure diversification phase, in which capital's initial entry into a landscape creates transient heterogeneity before eventually homogenizing it.

At $\alpha = 0$, the R\'{e}nyi entropy equals $\ln 7 = 1.946$~nats for all years, confirming that all seven classes are present throughout. The inter-year spread of $H_\alpha$ widens at high $\alpha$ (Figure~\ref{fig:entropy}b), indicating that year-to-year changes are concentrated in the dominant-class structure (Water and Built Area) rather than in the rare classes (Trees, Bare Ground). The largest KLD occurs during 2019--2020 ($D_{\mathrm{KL}} = 0.028$~nats), followed by 2022--2023 ($0.025$~nats), while the smallest value, $0.008$~nats (2021--2022), corresponds to a period of relative stability.

The FR distances on the 7-class simplex range from $0.119$~rad (2021--2022) to $0.261$~rad (2019--2020), with a mean of $\bar{d}_{\mathrm{FR}} = 0.166$~rad and $s_{\mathrm{FR}} = 0.050$~rad. The pulse threshold (Eq.~\ref{eq:pulse}) is $0.217$~rad. One transition clearly exceeds this threshold: 2019--2020 ($d_{\mathrm{FR}} = 0.261$~rad), corresponding to the opening and major construction phase of PIK2. The 2022--2023 transition ($d_{\mathrm{FR}} = 0.208$~rad) falls just below the threshold but represents the second-largest transformation and coincides with the pre-PSN land clearance spike identified in the simplex analysis. All three distance measures (KLD, FR, and Euclidean $d_{L_2}$) rank the seven periods consistently.

\begin{figure}[H]
    \centering
    \includegraphics[width=\linewidth]{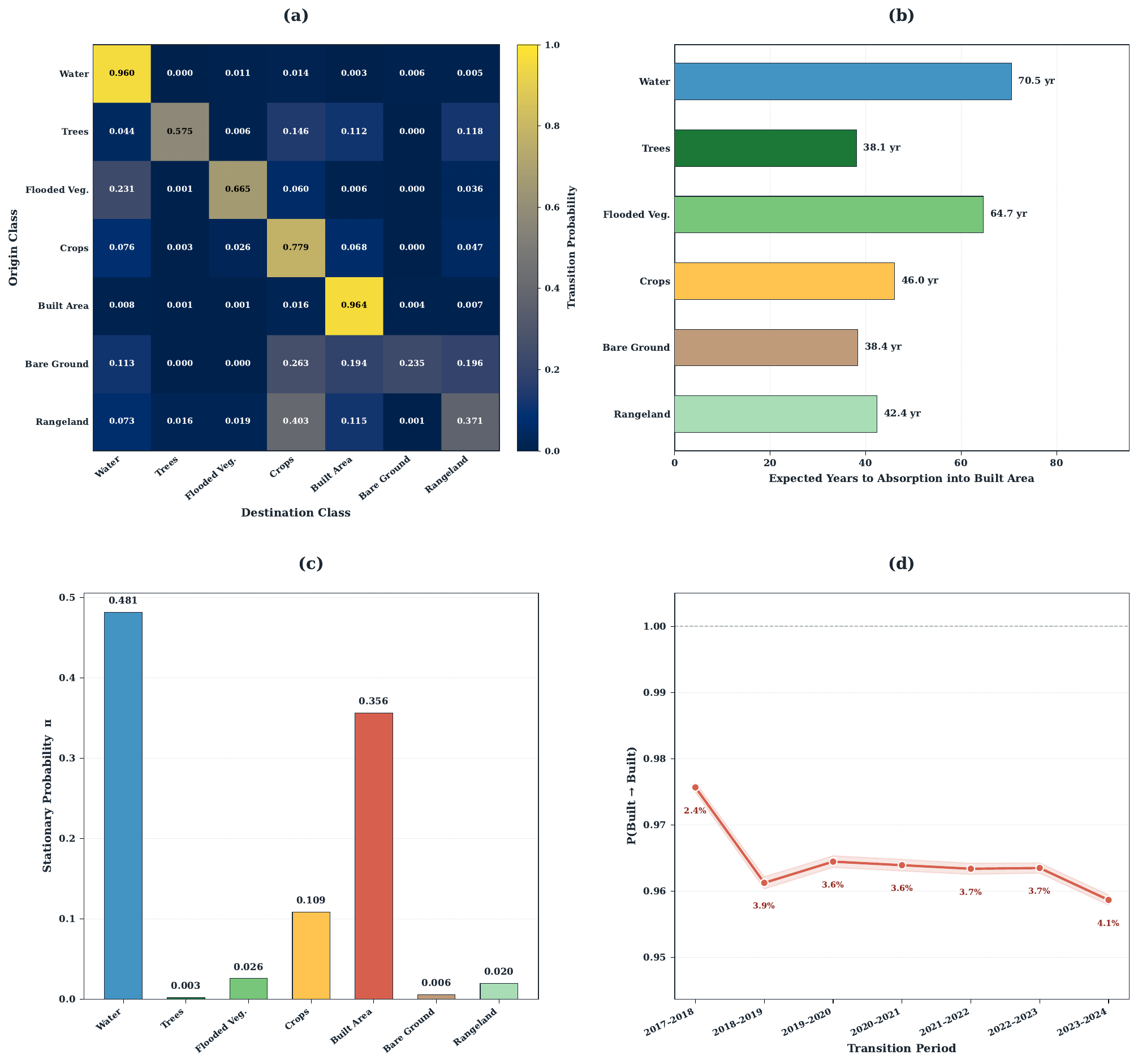}
    \caption{Markov chain analysis. (a) Pooled transition probability matrix. (b) Expected absorption time $\mathbb{E}[T_i]$ into Built Area under the absorbing chain formulation. (c) Stationary distribution $\boldsymbol{\pi}$ of the empirical (non-forced) pooled matrix. (d) Built Area self-retention probability $P(B \to B)$ by period, with leakage percentage annotated.}
    \label{fig:markov}
\end{figure}

Figure~\ref{fig:markov} presents the Markov chain analysis. The pooled count matrix aggregates $12{,}138{,}378$ pixel-pair transitions across seven consecutive year-pairs. The pooled matrix (Figure~\ref{fig:markov}a) reveals strong self-retention for Water ($P_{WW} = 0.960$), Built Area ($P_{BB} = 0.964$), and Crops ($P_{CC} = 0.779$). Bare Ground is the most volatile class ($P_{\mathrm{BG},\mathrm{BG}} = 0.235$), with substantial transition probabilities to Crops ($0.263$), Built Area ($0.194$), and Rangeland ($0.196$), confirming its role as construction-phase transitional land and validating its aggregation within the Capital category. Two structural zeros appear in the pooled matrix: Trees $\to$ Bare Ground and Flooded Vegetation $\to$ Bare Ground (zero observed transitions across $12.1$ million pairs).

The pooled matrix has leading eigenvalue $\lambda_1 = 1.000$, second eigenvalue $|\lambda_2| = 0.960$, spectral gap $1 - |\lambda_2| = 0.040$, and implied mixing time $t_{\mathrm{mix}} = -1/\ln|\lambda_2| \approx 24.5$ steps. The narrow spectral gap reflects the strong self-retention of Water and Built Area, which slow convergence to the stationary distribution $\boldsymbol{\pi}$. That distribution is $\pi_{\mathrm{Water}} = 0.481$, $\pi_{\mathrm{Built}} = 0.356$, $\pi_{\mathrm{Crops}} = 0.109$, with all other classes below $3\%$ (Figure~\ref{fig:markov}c). If the current transition dynamics were to persist indefinitely, over one-third of the landscape would converge to Built Area, while tree cover would shrink to $0.25\%$.

Under the absorbing formulation with $B$ as the absorbing state, the expected absorption times from the pooled transition matrix are: Trees, $\mathbb{E}[T] = 38.1$~yr ($\mathrm{SD} = 57.5$); Bare Ground, $38.4$~yr ($58.2$); Rangeland, $42.4$~yr ($59.8$); Crops (Agrarian), $46.0$~yr ($61.1$); Flooded Vegetation, $64.7$~yr ($65.6$); and Water, $70.5$~yr ($66.2$). The fact that $\mathrm{SD} > \mathbb{E}[T]$ for all classes reflects the heavy right tail of the first-passage-time distribution: most pixels are absorbed relatively quickly, but a fraction persists for much longer due to high self-retention in transient states. These standard deviations characterize the intrinsic stochastic spread of individual pixel trajectories, not estimation uncertainty (the expected values are exact given the transition matrix). Trees and Bare Ground face the shortest expected lifespans (${\sim}38$~yr), while Crops, the class most directly associated with smallholder agriculture, has an expected lifespan of $46.0$~yr. Flooded Vegetation and Water, buffered by their distance from the built frontier, persist longer ($65$--$71$~yr).

The period-specific self-retention rate $P(B \to B)$ oscillates narrowly between $0.959$ and $0.976$ (Figure~\ref{fig:markov}d). The corresponding leakage ($1 - P_{BB}$) of $2.4\%$--$4.1\%$ represents transitions predominantly between Built Area and Bare Ground (construction churn), not a return to agricultural or natural land uses. This quasi-absorbing behavior supports the interpretation that once land enters the built environment, it is structurally locked within the accumulation circuit.

The $G$-test decisively rejects the null hypothesis of temporal homogeneity: $G = 583{,}219$ with $\mathrm{df} = 252$ ($p < 10^{-10}$). The Frobenius norm $\|\mathbf{P}(t) - \mathbf{P}^{(\mathrm{pool})}\|_F$ ranges from $0.41$ to $0.92$ (mean $0.61$), confirming that the transition dynamics shift substantially from year to year. This non-stationarity is consistent with the political volatility surrounding PIK2: the largest Frobenius deviations occur in the early periods (2017--2018, $\|\cdot\|_F = 0.92$) and around the pre-PSN surge (2019--2020, $0.70$), reflecting discrete legislative and construction shocks rather than a smooth diffusive process \cite{Alfiana2026}.

\begin{figure}[H]
    \centering
    \includegraphics[width=\linewidth]{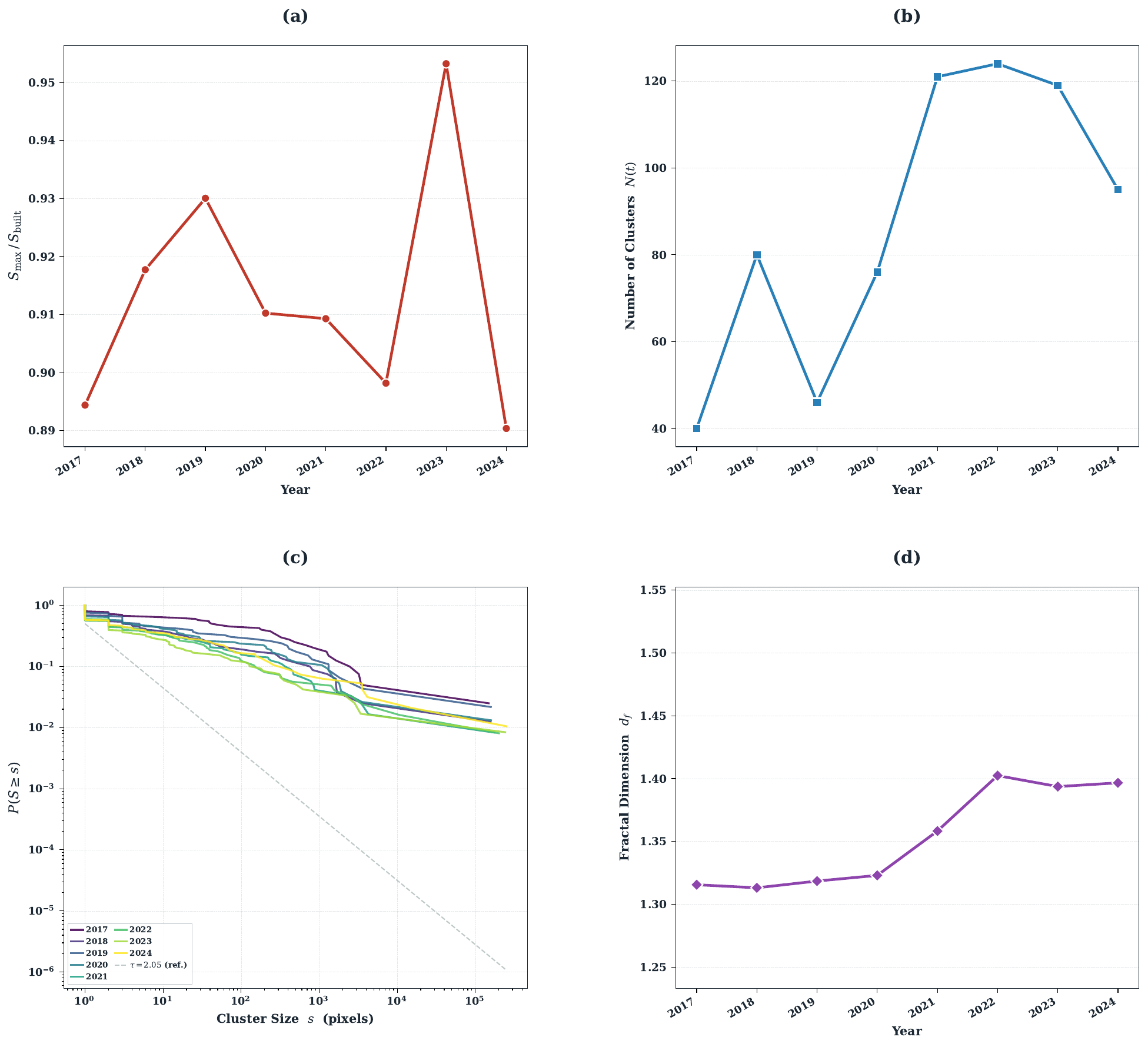}
    \caption{Percolation analysis of the built environment. (a) Order parameter $\Omega(t) = S_{\max}/S_{\mathrm{built}}$. (b) Number of connected clusters $N(t)$. (c) Complementary cumulative distribution function (CCDF) of cluster sizes. (d) Box-counting fractal dimension $d_f$ of the largest cluster boundary.}
    \label{fig:percolation}
\end{figure}

Figure~\ref{fig:percolation} summarizes the percolation analysis. The occupation probability $p(t)$ increases from $0.096$ (2017) to $0.162$ (2024), remaining far below $p_c \approx 0.593$ throughout. Despite this, the order parameter $\Omega(t)$ consistently exceeds $0.89$, reaching a peak of $0.953$ in 2023 (Figure~\ref{fig:percolation}a). A single giant connected component thus contains $89\%$--$95\%$ of all built pixels. Under uncorrelated random site occupation, achieving $\Omega > 0.9$ at $p \approx 0.1$ would be extraordinarily unlikely. The observed supercriticality is a signature of strong spatial correlations introduced by planned road networks and infrastructure corridors, which create long-range connectivity at low global occupation densities.

The number of connected clusters increases from $40$ (2017) to a peak of $124$ (2022), then decreases to $95$ (2024) (Figure~\ref{fig:percolation}b). The simultaneous increase in both cluster count and $\Omega$ during 2017--2022 indicates that new satellite developments are appearing at the urban frontier while the core giant component consolidates and grows. The subsequent decline in cluster count by 2024 suggests that some peripheral developments have merged into the giant component.

The cluster-size distribution is extremely heavy-tailed: in 2024, the largest cluster ($S_{\max} = 249{,}594$ pixels) is over 16 times the size of the second-largest ($15{,}009$), and the median cluster size is just 2 pixels. The fraction of isolated single-pixel clusters increases from $22\%$ (2017) to $43\%$ (2024).

The box-counting fractal dimension of the giant component's boundary increases from $d_f = 1.316$ ($R^2 = 0.993$) in 2017 to $d_f = 1.397$ ($R^2 = 0.993$) in 2024, with a mean of $1.353$ across all years (Figure~\ref{fig:percolation}d). These values place the urban boundary morphology between compact growth ($d_f \to 1$) and Eden-model growth ($d_f \approx 1.5$), below the DLA regime ($d_f \approx 1.71$). The secular upward trend suggests increasing frontier irregularity as the mega-development extends into the less-planned agrarian periphery. The rising $d_f$ is consistent with a transition from road-led, corridor-based expansion (relatively compact) toward more irregular, tentacle-like penetration into surrounding smallholder land, a pattern previously associated with speculative urban growth in other developing-world megacities \cite{Makse1995, Batty2007}.

\section{Discussion}
\label{sec:discussion}

The information-theoretic analysis reveals that the landscape is currently in a pre-enclosure diversification phase. Entropy is increasing because the growth of Built Area ($9.6\% \to 16.2\%$) partially redistributes probability mass away from the Water-dominated baseline, creating transient heterogeneity. This finding cautions against interpreting rising entropy as evidence of ecological health or spatial balance; in this context, it reflects the initial fragmentation that accompanies capital's entry into a new frontier. Shannon entropy alone is an insufficient diagnostic for enclosure and must be interpreted jointly with the class-specific probability trajectories and the simplex analysis.

The FR distance on the full 7-class simplex identifies the 2019--2020 transition as the dominant transformation pulse. On the reduced 3-class simplex, this same period records the highest velocity ($0.405$~rad/yr). The alignment of these two independently computed metrics lends confidence to the identification of 2019--2020 as the primary construction-phase discontinuity. The 2022--2023 period, while not exceeding the formal pulse threshold on the 7-class simplex, registers as the second-largest transformation on both manifolds and coincides with the documented spike in bare-ground clearance preceding the 2024 PSN designation \cite{Kemenko2024PSN}.

The absorbing Markov chain analysis quantifies the temporal horizon of enclosure. The expected absorption time for Crops (Agrarian) is $46.0$ years under pooled transition dynamics. This figure should be interpreted with caution: it assumes stationarity of the transition matrix, which the $G$-test decisively rejects. If the recent acceleration in construction activity (visible in the declining period-specific absorption times, particularly in 2022--2023 where $\mathbb{E}[T_{\mathrm{Crops}}]$ falls to $10.9$~yr) is sustained, actual absorption could occur substantially faster. Conversely, the political pushback against PIK2 following the 2025 review of PSN status \cite{Alfiana2026} could slow or partially halt the process.

The percolation results are perhaps the most diagnostic for distinguishing planned mega-development from organic urban growth. The observed supercriticality ($\Omega > 0.89$) at occupation probabilities ($p \approx 0.10$--$0.16$) roughly one-sixth of $p_c$ would be essentially impossible under uncorrelated site percolation. It provides quantitative evidence that the spatial structure of PIK2 is determined by planned infrastructure (road grids, utility corridors) that introduce long-range spatial correlations, creating a giant connected component at very low occupation densities \cite{Makse1995, Rozenfeld2008}. This distinguishes PIK2 from the organic, incrementally accreting settlement patterns typical of informal urbanization.

The social implications of these findings warrant consideration. The nearly one-to-one Agrarian-to-Capital transfer ($-13.2$~pp crops vs.\ $+14.1$~pp capital) quantifies the displacement of smallholder agricultural land. For the approximately $541{,}000$ land pixels ($54.1$~km$^2$) classified in 2024, Agrarian constitutes $29.3\%$ (${\sim}15.9$~km$^2$), a substantially reduced agricultural base compared to 2017 ($42.5\%$, ${\sim}18.1$~km$^2$). The communities dependent on these agricultural lands, predominantly rice paddies and aquaculture ponds on the peri-urban coastal fringe, face livelihood displacement that the aggregate statistics only partially capture.

The finding that speculative land clearance (the 2023 Bare Ground spike) preceded the formal PSN designation raises questions about the sequencing of regulatory authorization and private-sector land preparation, a pattern that has been documented in reporting on agrarian conflicts in the PIK2 corridor \cite{Alfiana2026}. The quantitative findings of our spatial models provide a physical baseline that corroborates these recent socio-legal critiques of the PIK2 mega-development. Public policy analyses indicate that the project's rapid enclosure, which our percolation analysis identifies as supercritical and structurally planned, was facilitated by top-down spatial planning that systematically neglected local community participation \cite{Latifah2024, Fauziah2025}. The dominant Agrarian-to-Capital transfer we observed translates directly to the structural displacement of traditional fishers and farmers in coastal villages such as Muara, Kronjo, Tanjung Pasir, and Kohod \cite{Alfiana2026}. This physical erasure of agrarian space has severe socio-economic and ecological consequences; for instance, the destruction of vital mangrove ecosystems and restricted marine access have caused up to 65\% of local fishermen to experience significant income declines \cite{Latifah2024, Fauziah2025}.

Furthermore, the non-stationary Markov transition dynamics and the pre-PSN bare-ground clearance spike we identified reflect a highly volatile and contested regulatory environment. The contradiction between the project's branding as an eco-friendly "Tropical Coastland" and the actual, irreversible expansion of capitalist infrastructure eventually triggered profound institutional pushback. Recently, the Supreme Court of Indonesia (Decision No. 12 P/HUM/2025) revoked the project's National Strategic Project (PSN) status, ruling that its regulatory foundation conflicted with higher environmental, spatial planning, and forestry laws \cite{Alfiana2026}. Our information-theoretic framework thus quantitatively captures the physical signature of this legal friction, demonstrating that the spatial evolution of PIK2 is not driven by organic urban diffusion, but rather by a legally contested pulse of primitive accumulation.

Several limitations should be noted. First, the Sentinel-2 LULC product, while globally consistent, relies on deep-learning classification that may introduce systematic errors, particularly in the distinction between Flooded Vegetation and Water in tidal zones. Second, the Markov chain model treats each pixel independently, ignoring spatial autocorrelation in transition probabilities. Third, the absorbing chain formulation imposes perfect irreversibility on Built Area, which is an idealization (the empirical leakage of $2$--$4\%$ is nonzero, though it represents construction churn rather than de-commodification). Fourth, our analysis is confined to the spatial domain of PIK2 and its immediate surroundings; we do not model displaced populations, economic flows, or the broader housing market dynamics that drive speculative development.

\section{Conclusion}
\label{sec:conclusion}

We have presented a quantitative framework combining IG, absorbing Markov chains, and percolation theory to characterize the spatial dynamics of land enclosure in the PIK2 mega-development near Jakarta, Indonesia. The FR geodesic distance isolates discrete construction pulses; the absorbing Markov chain quantifies expected absorption times for each LULC class; and percolation analysis reveals planned supercritical connectivity and an increasingly irregular fractal frontier. Together, these tools provide a statistical-mechanical diagnostic that can, in principle, be applied to any georeferenced LULC time series where the dynamics of spatial commodification are of interest. The framework bridges the quantitative apparatus of statistical physics with the structural categories of political economy, offering a path toward more precise empirical engagement with processes that have traditionally been analyzed through qualitative or purely descriptive methods.

\section*{Declaration of competing interest}
The authors declare that they have no known competing financial interests or personal relationships that could have appeared to influence the work reported in this article.

\section*{Declaration of generative AI use}
During the preparation of this work, the authors used Claude 4.6 Sonnet (Anthropic) and Gemini 3.1 Pro (Google) solely for the purposes of English grammar, vocabulary refinement, and improving the overall readability of the manuscript. The authors maintain full responsibility for the conceptualization, model development, computational execution, mathematical derivations, and the integration and analysis of social theory presented in this study.

\section*{Data availability}
The original Sentinel-2 LULC data are publicly accessible at \url{https://livingatlas.arcgis.com/landcoverexplorer}. The Python scripts are available at \url{https://github.com/sandyherho/supplPIK2LULC}. Processed datasets, figures, and statistical reports are hosted on the Open Science Framework (OSF) and can be accessed via \url{https://doi.org/10.17605/OSF.IO/ZTWVU}. All materials are released under the MIT License.

\section*{Acknowledgements}
Financial support was provided by the ITB 3P Research Program (Talenta Unggul Scheme) through the Directorate of Research and Innovation, Bandung Institute of Technology (Project ID: DRI.PN-6-64-2026).

\end{document}